\documentclass{elsart}
\usepackage{natbib}

\usepackage{graphics}

\newcommand{\FIG}[1]{}

\newcommand{\vv}{\mbox{\bf v} \,}
\newcommand{\xx}{\mbox{\bf x}}
\newcommand{\ggg}{\mbox{\bf g}}
\newcommand{\UU}{\mbox{\bf U}}
\newcommand{\FF}{\mbox{\bf F}}
\newcommand{\BB}{\mbox{\bf B} \,}
\newcommand{\SSS}{\mbox{\bf S}}
\newcommand{\divb}{\left(\nabla \cdot \BB\right)}
\newcommand{\ptot}{p_{\mbox{\small total}}}

\begin{document}
\runauthor{Keppens, Nool, T\'oth, Goedbloed}
\begin{frontmatter}
\title{Adaptive Mesh Refinement for conservative systems: multi-dimensional
efficiency evaluation}
\author[fom]{R. Keppens\thanksref{email}}
\author[cwi]{M. Nool}
\author[buda]{G. T\'oth}
\author[fom]{J.P. Goedbloed}

\address[fom]{FOM-Institute for Plasma Physics Rijnhuizen, 
Association Euratom-FOM, \hfill \\
P.O. Box 1207, 3430 BE Nieuwegein, The Netherlands}
\address[cwi]{Centre for Mathematics and Computer Science, P.O. Box 94079,
\hfill \\
1090 GB Amsterdam, The Netherlands}
\address[buda]{Department of Atomic Physics, E\"otv\"os University, \hfill \\
P\'azm\'any P\'eter s\'et\'any 1, 1117 Budapest, Hungary}
\thanks[email]{Corresponding author: keppens@rijnh.nl, tel: (+31) 30-6069941,
fax: (+31) 30-6031204}

\begin{abstract}
Obtainable computational efficiency is evaluated when
using an Adaptive Mesh Refinement (AMR) strategy in time accurate simulations
governed by sets of conservation laws.
For a variety of 1D, 2D, and 3D
hydro- and magnetohydrodynamic simulations, AMR is used in combination with
several shock-capturing, conservative discretization schemes. Solution
accuracy and execution
times are compared with static grid simulations at the corresponding high
resolution and time spent on AMR overhead is reported.
Our examples reach corresponding
efficiencies of 5 to 20 in multidimensional calculations
and only 1.5 -- 8 \% overhead is observed.
For AMR calculations of multi-dimensional magnetohydrodynamic problems,
several strategies for controlling the $\nabla \cdot \BB =0$ constraint
are examined. Three source term approaches suitable for cell-centered
$\BB$ representations are shown to be effective. For 2D and 3D calculations
where a transition to a more globally turbulent state takes place,
it is advocated to use an approximate Riemann solver based discretization
at the highest allowed level(s), in combination with the robust
Total Variation Diminishing Lax-Friedrichs method on the coarser levels.
This level-dependent use of the spatial discretization acts as a
computationally efficient, hybrid scheme.
\end{abstract}

\begin{keyword}
PACS 52.30.-q; 02.70.Fj; 02.30.Jr \hfill \\
Conservation Laws -- Magnetohydrodynamics -- Adaptive Mesh Refinement -- Shock-capturing -- Computational Efficiency
\end{keyword}
\end{frontmatter}

\typeout{SET RUN AUTHOR to \@runauthor}

\section{Introduction}\label{s-focus}

Adaptive Mesh Refinement (AMR), an $h$-refinement method for solution-adaptive
computations, is used in an ever increasing number of plasma physical and
astrophysical simulations. 
The AMR algorithm and, specifically,
all complications associated with ensuring global conservation properties
when dealing with hyperbolic conservative systems, 
have been addressed in 
the work by Berger and Colella~\cite{berger}. In that pioneering
paper, the authors stress that 
{\it ``\ldots it seems possible to implement a general
code where the number of dimensions is input''.} We make use of such a
general code in all simulations presented here, where both the dimensionality
and the particular system of conservation laws is selected in a pre-processing
stage. The resulting versatility in the applications allows us to assess 
quantitatively how much can be gained from employing AMR 
in 1D, 2D, or 3D hydro- and magnetohydrodynamic (MHD) simulations, for a variety
of spatial discretizations. 
Our coding effort benefits from the expertise gained
through many general purpose and 
problem-tailored AMR software packages, which  
have been in active use and continued development for more than a decade. 
E.g., AMRCLAW~\cite{webamrclaw} by LeVeque and
coworkers combines AMR with a selection of time-dependent
hyperbolic systems of partial differential equations, including both
conservative and nonconservative systems. Berger and LeVeque~\cite{amrclaw}
recently demonstrated the use of multi-dimensional wave propagation
algorithms~\cite{leveque,lang} in 2D adaptive computations on Cartesian and
logically rectangular, curvilinear grids. 
A related development effort by Walder and Folini~\cite{amaze,walder2} 
is AMRCART, a software package
which intends to encompass 3D adaptive MHD scenarios.
Dimension-independent library approaches for AMR simulations include 
the publicly available PARAMESH toolkit by 
MacNeice {\it et al.}~\cite{paramesh},
as well as {\tt C++} class framework approaches such as the
{\it AmrLib} extension to the {\it BoxLib} library~\cite{lbnl}. 
The code used here is the most recent extension to the Versatile Advection
Code (see {\tt http://www.phys.uu.nl/}$\sim${\tt toth}), 
briefly discussed in section~\ref{ss-vac}.

Our AMR implementation differs somewhat from the original algorithm 
in~\cite{berger} and these differences will be mentioned in 
section~\ref{ss-amr}. 
For several multidimensional calculations, 
we report on computational efficiency and solution accuracy.
We beneficially combine a fully upwind scheme on the
highest AMR level with a Total Variation Diminishing Lax-Friedrichs
discretization on all other levels.
For 2D hydrodynamic and magnetohydrodynamic problems, 
we thereby typically gain a factor of 10 in execution time
as compared to the equivalent high resolution static grid result. 
Our example simulations cover 1D Riemann problems (Section~\ref{ss-res1d}),
a 2D advection problem (Section~\ref{sss-vac2d}), 
and both shock-dominated and more turbulently
evolving hydro- and magnetohydrodynamic multidimensional cases. 
Tables~\ref{t-time}-\ref{t-time3d} collect
our findings as presented in Section~\ref{s-res}.

As our ultimate aim is to perform grid-adaptive,
realistic 3D MHD simulations to investigate general
plasma physical and astrophysical scenarios, we pay particular attention
to strategies for controlling the solenoidal character of the magnetic
field. In section~\ref{ss-divb},
related efforts in AMR MHD simulations are discussed
in relation to their specific treatment of this constraint.
Recently, Balsara~\cite{balsara} and independently
T\'oth and Roe~\cite{toth-roe} introduced a divergence-free 
reconstruction and prolongation strategy for multi-dimensional AMR MHD
simulations which involves staggered magnetic field components. In this paper,
we evaluate three alternative,
simple approaches with cell-centered $\BB$ components 
where $\nabla \cdot \BB=0$ 
is maintained to truncation error by the addition of suitable source terms.
Quantitative comparisons between AMR MHD calculations and high resolution
static grid runs in section~\ref{ss-res2d}
demonstrate their effectiveness.

\section{Algorithm}\label{s-amrvac}

\subsection{Versatile Advection Code, equations and solvers}\label{ss-vac}

The Versatile Advection Code (VAC), 
initiated by T\'oth~\cite{vac1}, is specifically
designed for simulating dynamics governed by a system of (near-)
conservation laws. 
It's versatility resides in the choices offered in the
geometry of the grid, the dimensionality, the physical application
(including Euler and MHD equation modules), the
spatial~\cite{vac2} and temporal discretization~\cite{impl1,impl2},
and the computer platform~\cite{vacpar}.

VAC uses various second order shock-capturing numerical algorithms:
the Flux Corrected Transport (FCT) scheme \cite{boris},
the Lax-Wendroff type Total Variation Diminishing (TVD) \cite{harten}
and the TVD-MUSCL \cite{vanleer} schemes (TVDMU)
with Roe-type approximate Riemann solvers \cite{roe, roe-balsara},
and the TVD Lax-Friedrichs (TVDLF) method \cite{yee}.
For the TVD type schemes different slope limiters are available, including
the most robust {\it minmod}, and the sharper {\it monotonized central}
(MC, also referred to as Woodward) limiter.
For exact specifications of the algorithms, see~\cite{vac2}
and references therein. 

In multi-dimensional MHD simulations one needs to control
the numerical value of $\nabla \cdot \BB$. VAC provides
several algorithms, such as the eight-wave scheme~\cite{powell},
the projection scheme~\cite{brack}, 
and several types of constrained transport and central
difference type discretizations, which were recently described and evaluated for
static grid simulations by T\'oth~\cite{divb}. In section~\ref{ss-divb}, we
discuss three strategies which easily carry over to an adaptive grid. 
Until now, VAC had no automated means for dynamic grid adaptation.
Our new AMRVAC software package combines the versatility of VAC with
the advantages of adaptive mesh refinement. In this first implementation,
only Cartesian grids and explicit time stepping is allowed.

We demonstrate the flexibility of AMRVAC
by using three equation modules in 1, 2 and 3 dimensions with
different conservative high resolution spatial discretization schemes.
In the example simulations from Section~\ref{s-res}, the equation modules
are subsets of
the system of magnetohydrodynamics, given by
\begin{eqnarray}
\frac{\partial \rho}{\partial t}+\nabla\cdot(\vv\rho) & = & S_{\rho}
                                                       \label{q-rho} \\
\frac{\partial \rho \vv}{\partial t}
+\nabla\cdot(\vv\rho\vv-\BB\BB) +\nabla \ptot & = & \SSS_{\rho v}
                                                       \label{q-momentum} \\
\frac{\partial e}{\partial t}
+\nabla\cdot(\vv e+\vv \ptot-\BB\BB \cdot \vv) & = & S_e
                                                       \label{q-energy} \\
\frac{\partial\BB}{\partial t}+\nabla\cdot(\vv\BB-\BB\vv)& = & \SSS_B.
                                                       \label{q-induction}
\end{eqnarray}
These are expressed in the conservative variables 
density $\rho$, momentum density $\rho\vv$,
total energy density $e$, and magnetic field $\BB$, with all additional
sources and sinks collected in the source terms $S_\rho$, $\SSS_{\rho v}$, $S_e$
and $\SSS_B$. In these equations, the total kinetic and magnetic pressure is
written as
\begin{equation}
\ptot=\left(\gamma - 1\right)\left(e-\frac12\rho\vv^{2}-\frac12\BB^{2}\right)+\frac12\BB^{2},
\end{equation}
for an ideal gas with adiabatic index $\gamma$. We stress that for all
simulations done, we time advance the relevant subset in the selected
dimensionality only, which is chosen {\em prior} to compilation of the code.
This is possible because
the entire AMR algorithm is implemented in the 
LASY-syntax~\cite{lasy}, making it effectively dimension-independent.
Similar to VAC, the AMRVAC package is configured to dimensionality
using the VAC preprocessor (a Perl script), prior to compilation.

The time step is calculated to comply
with the Courant-Friedrichs-Levy (CFL) stability criterion. If not stated
otherwise, the Courant number is 0.9. The 
conservative high resolution spatial discretization schemes used are
FCT, TVDLF, TVD, and TVDMU. 
These shock-capturing schemes differ in the numerical representation of the
fluxes $\FF_i$ in the set of equations written as
\begin{equation}
  \partial_t\UU + \sum_i\partial_i\FF_i(\UU) =
     \SSS (\UU,\partial_i \UU, \partial_i\partial_j \UU, \xx, t).
\label{q-system}
\end{equation}
Here, the conservative
variables are collected in $\UU(\xx, t)$, and $i$ and $j$
run over 1, 2 or 3 components of the spatial coordinate $\xx$.
If not stated otherwise, multidimensional scenarios add fluxes from the
different orthogonal directions simultaneously, in contrast to a dimensionally
split~\cite{strang} approach. 
The source terms $\SSS$ can also be unsplit, i.e. added at the same time as the
fluxes, or split, i.e. added before and after the update with fluxes.

\subsection{Adaptive Mesh Refinement}\label{ss-amr}

Detailed descriptions of the AMR algorithm can be found in Berger
and Colella~\cite{berger}, Bell {\it et al.}~\cite{bell}, 
Friedel {\it et al.}~\cite{grauer}, Steiner {\it et al.}~\cite{oski}, 
and many others. We summarize the essential steps and its
user controlled features. Thereby, we pay specific attention
to details which differ from the original 
Berger and Colella~\cite{berger} approach. 

\subsubsection{Time stepping and conservation}\label{sss-amr1}

In essence, AMR represents an automated procedure to generate or
destruct --- both controlled by the ensuing dynamics -- hierarchically
nested grids with subsequently finer mesh spacings $\Delta x_i$ (index $i=1,\ldots,D$ for a $D$-dimensional problem).
Up to a predefined maximal grid
level $l_{\rm max}$, consecutive levels $l$ are characterized by 
even refinement ratios $r_l$ with $l\in [2,l_{\rm max}]$, such that
\begin{equation}
r_l \equiv \frac{\Delta x_{i,l-1}}{\Delta x_{i,l}} \equiv 
\frac{\Delta t_{l-1}}{\Delta t_{l}}.
\end{equation}
To make the global conservation property computationally tractable, we stick to
Cartesian grids and the 
`proper nesting' criterion~\cite{berger}: (i)
level $l>1$
grid boundaries coincide with grid lines of $l-1$ meshes; and (ii) we insist on
the telescoping hierarchy of $l-2$, $l-1$ and level $l$ grids, except
at (non-periodic) physical boundaries. 

Consequently, two -- related -- issues for an AMR simulation are (1)~how one
time-advances on such a sequence of nested and thus overlapping grids, while
(2)~keeping the solutions consistent and conservative through a suitable
`update' process. 
The approach taken is
illustrated in Figure~\ref{f-time}, showing a 
hypothetical sequence of three time steps in a case where $l_{\rm max}=4$.
The schema is traversed from left to right, bottom to top, and with
horizontal `update' arrows preceeding vertical time `advance' steps. Each time
a level $l>1$ has caught up in time with the underlying $l-1$ grids, all level
$l-1$ cells that are covered by the finer grids are replaced by conservative
averages. With notation defined in Figure~\ref{f-update} for a 2D example where
$r_l=2$, this is achieved by simple averaging since e.g. the density in
the underlying coarse cell must be $\rho_{ij}\Delta x_{1,l-1} \Delta x_{2,l-1} =
\sum_m \sum_n \rho_{mn} \Delta x_{1,l} \Delta x_{2,l}$ so that
$\rho_{ij}= \sum_m \sum_n \rho_{mn} / r_l^2$. 
Also, those $l-1$ cells 
immediately adjacent to a level $l$ grid
boundary are modified to comply with the interface fluxes used on the 
finer level. This step is crucial for restoring the global conservation
property across the AMR grid hierarchy. 
Following the notation used in Figure~\ref{f-update}, this involves a
change in the temporal update for the density in coarse cell indexed
by $ij$ on level $l-1$ bordered on the right hand side by level $l$ cells.
While originally
\[
\rho_{ij}^{(n+1)} = \rho_{ij}^{(n)} - \frac{\Delta t}{\Delta x_{i,l-1}}\left(
F_{i+\frac{1}{2} j}^{(n+\frac{1}{2})}- F_{i-\frac{1}{2} j}^{(n+\frac{1}{2})} \right)
- \frac{\Delta t}{\Delta x_{j,l-1}}\left(
F_{i j+\frac{1}{2}}^{(n+\frac{1}{2})}- F_{i j-\frac{1}{2}}^{(n+\frac{1}{2})} \right)
,
\]
the time centered numerical flux across the right cell edge 
$F_{i+\frac{1}{2} j}^{(n+\frac{1}{2})}$
needs to be replaced by
\[
F_{i+\frac{1}{2} j}^{(n+\frac{1}{2})} \Rightarrow \frac{1}{r_l^2}\left[
F_{m-\frac{1}{2} n}^{(n+\frac{1}{4})} +
F_{m-\frac{1}{2} n+1}^{(n+\frac{1}{4})} +
F_{m-\frac{1}{2} n}^{(n+\frac{3}{4})} +
F_{m-\frac{1}{2} n+1}^{(n+\frac{3}{4})}\right]. 
\]
This flux fix restores the conservation.
For more complete discussions of these update and flux fixing steps, 
we refer to~\cite{berger}.

In our AMR implementation, each level solves the same set of 
equations but can exploit a different 
discretization scheme per level. This in contrast
with the more general Adaptive Mesh and Algorithm Refinement (AMAR)
approach introduced by Garcia {\it et al.}~\cite{garcia}, who coupled
a continuum description with a particle method on the finest level
of the AMR hierarchy. AMAR is
possible because the only communication between different level solutions
is through (i) the update process and (ii) the filling of
`ghost' cells as used for imposing boundary conditions. 

The time advancing is further
complicated by the automated regridding operation, as indicated by
the grey circles in Figure~\ref{f-time}.
This regrid action is controlled by a simple criterion:
when at least $k$ time steps are taken on a certain level, this level is
evaluated for refinement ($k=2$ in the figure). This affects all
higher level grids which may either suddenly appear
(in Figure~\ref{f-time} after the first and halfway in the second time step),
disappear (after the second time step), or simply get rearranged or be
left unchanged (halfway in the third time step). Of course, all level
$1$ grids remain unaltered, while the maximal allowed
level $l_{\rm max}$ is never evaluated for further refinement.

\subsubsection{Refinement criteria and regridding}\label{sss-amr2}

Conceptually, regridding consists of two steps: in a first
step, cells on level $l$
which need to be refined are identified - `flagged' - through physically and
numerically controlled criteria. The second step ensures that a properly
nested new grid level $l+1$ is formed which encompasses most, preferably all,
of the flagged cells. 
As pointed
out above, the entire grid tree above a certain fixed base level $l_{\rm base}$
may change in that process. With the current finest level in existence
being $l_{\rm fine}$, one descends in levels from $l=\min (l_{\rm fine},
l_{\rm max}-1)$ down to $l=l_{\rm base}$.

In the first step -- the flagging procedure -- a 
great variety of refinement criteria can be adopted. 
Powell {\it et al.}~\cite{powel} use local measures of compressibility, 
rotationality, and current density in their 3D MHD simulations and a region is
refined when chosen treshold values are exceeded. Ziegler~\cite{ziegler} uses
similarly motivated, though more abstract, `gradient value based' criteria
where the slopes of functionals of the fluid variables, 
like the sound speed $c_s$, control refinement. In cases where 
an (approximate) Riemann solver based method is exploited, 
one may flag cells in a `wave-affected region'. The extent
of this region can be computed
from the wave strengths and speeds as calculated in the Riemann solver.
Such criterion was succesfully exploited for unsteady Euler 
flows in~\cite{chiang}.

Our refinement criterion to `flag' cells is a variant of 
the error estimation procedure from Berger and Colella~\cite{berger},
which is generally applicable to any system of equations and integration
method used. The idea is to rely on Richardson extrapolation, by comparing 
a solution for a level $l$ obtained by coarsening and integrating with
one which is integrated first, and subsequently coarsened. 
Bell {\it et al.}~\cite{bell} improved on this basic idea by additionally
exploiting the uncoarsened solution vector as well, to avoid missing
features that get lost in the averaging. Moreover, they allowed for
a certain amount of user-forced refinement or derefinement. 
In our implementation, the 
integration method in the error estimation
can be selected as any one-step, dimensionally unsplit method, 
in combination with source terms. The
criterion to flag cells can also exploit uncoarsened, high-resolution
data, since we keep two full solution vectors
$\UU_l^{n-1}$ and $\UU_l^{n}$ on each level $l<l_{\rm max}$, 
separated in time by the corresponding $\Delta t_l^{n-1}$. For a fair
variety of test problems (including the examples described below and those
in Nool and Keppens~\cite{nool}, Keppens {\it et al.}~\cite{parcfd,iccs02}),
we found it sufficient to employ a first order
version of the TVDLF method in the error estimation procedure, while
integrating with a level-dependent mixture of TVD, TVDLF, or TVDMU on 
the full AMR hierarchy. The most crucial matter is to select the
right combination of physics variables and mutual weighting factors to
flag for. This is done 
as soon as a critical tolerance level $\epsilon_{\rm tol}$ is
exceeded. In the examples from section~\ref{s-res}, we always list
the chosen variables and tolerances.

Given any procedure to identify cells which need refining, the AMR scheme 
must ultimately arrange them into properly nested `rectangular' grids. 
Note that `rectangle' is to be interpreted according to dimensionality.
The basic steps to create this grid hierarchy were laid out in Berger and
Colella~\cite{berger}. In the original algorithm, each flagged cell was
surrounded with a buffer zone of width $n_{\rm buff}$ cells
in each direction. The collection of
flagged cells was subsequently made compatible with the nesting
criterion. Some of the subtleties involved with ensuring
this proper nesting are illustrated in Figure~\ref{f-grid}. Eventually,
the collection of properly nested level $l$ cells from which to
form the level $l+1$ grids need to be processed into `rectangles'. 
This latter step
was based on a succession of {\it bisections} of 
the encompassing `rectangle', followed by 
a {\em merge} step. 
Bisections were repeated till all grids
had their ratio of flagged to total points within the rectangle greater
than a predetermined efficiency ${\rm E} \in [0,1]$. 
The merging was meant to balance the possible creation of
many small grids. 
In its original form, this clustering algorithm
created acceptable, but non-optimal new candidate grids which would
often overlap and could end up with efficiencies below
${\rm E}$, due to the merging process.
Berger and Rigoutsos~\cite{rigout} designed a more sophisticated algorithm
to overcome these problems, where the grid partitioning was done
along edges where transitions from flagged to unflagged regions
occured. The edges were detected using pattern recognition techniques
exploiting signature arrays. This clustering algorithm has been used
on realistic 3D problems by Bell {\it et al.}~\cite{bell}, 
Balsara~\cite{balsara}, and various others. We chose to modify the
original merge step from~\cite{berger} to avoid the
creation of overlapping grids at the same nesting level. At the same time,
we introduced an enforced minimal
efficiency ${\rm E}_{\rm min} \in [0,1]$
on all resulting new grids. It could be of
interest to study how our modifications of the clustering algorithm
compare with the Berger and Rigoutsos~\cite{rigout} approach on
a variety of 2D and 3D synthetic input clusters.
In all simulations reported in what follows, we took ${\rm E}_{\rm min}=0.5$.

The resulting new grid structure is then initialized from the solutions
available on levels $l_{\rm base}$ and up. For those cells which can not
be copied directly from the same level on the original grid hierarchy, we use a
conservative, {\em minmod}
limited linear interpolation from the coarser level underneath. Due to the
nesting property, this is always possible. In a notation similar to the one used
in Figure~\ref{f-update}, this limited linear interpolation in a 2D case
would read
\[ \rho_{mn} = \rho_{ij} + \frac{x_{1,m} - x_{1,i}}{\Delta x_{1,i}} \Delta_i \rho^{(j)} + \frac{x_{2,m} - x_{2,i}}{\Delta x_{2,i}} \Delta_j \rho^{(i)} \]
where $\Delta_i \rho^{(j)}$ denotes the minmod limited slope in direction $i$
involving left-sided, right-sided, and centered slopes determined
from $[\rho_{i-1 j}, \rho_{ij}, \rho_{i+1 j}]$. Note that this is a conservative
operation and second order accurate in smoothly varying density regions.

\subsection{Multi-dimensional AMR MHD: the $\nabla \cdot \BB=0$ constraint}\label{ss-divb}

Multidimensional MHD computations that make use of 
AMR have only been emerging over the
last few years. Such computations must handle
the additional complication of maintaining a solenoidal magnetic field.
Two-dimensional adaptive calculations for incompressible MHD
flows were presented by Friedel {\it et al.}~\cite{grauer}. 
Those authors borrowed 
projection method techniques from incompressible Navier-Stokes 
simulations~\cite{bell2}. Details of these projection methods and how
to handle them on adaptive grid hierarchies have been discussed in
Almgren {\it et al.}~\cite{almgren}.

Steiner {\it et al.}~\cite{oski} performed
compressible MHD calculations using Flux Corrected Transport~\cite{boris}
in two space dimensions. They combined AMR with a staggered
representation where the magnetic field components are defined at 
cell interfaces while all other conserved quantities are cell-centered.
Balsara~\cite{balsara} has presented and applied
a divergence-free AMR MHD scheme where the same staggering was used.
The key elements in his scheme are a divergence free prolongation
strategy to transfer coarse solution information to finer grids, and a
divergence free restriction step to update coarse grids from overlying
finer levels. Similar to the flux fix-up step inherent in the
Berger and Colella~\cite{berger} method to ensure the global
conservation property, a similar electric field correction
step is needed at mesh level interfaces to keep the overall solutions
consistent. 

Powell {\it et al.}~\cite{powel} presented full
3D ideal MHD scenarios where solution-adaptive refinement is combined
with an eight-wave Roe-type approximate Riemann solver method. Representative
calculations of the magnetized solar wind impacting on a
planetary magnetosphere have been reported by those authors.
They advocated the use of
an eight-wave formulation, which maintains the
$\nabla \cdot \BB=0$ constraint to truncation error. The approach
modifies the approximate Riemann solver 
to `propagate' $\nabla \cdot \BB$
errors with the plasma velocity and adds non-conservative source
terms to the MHD equations proportional to $\nabla \cdot \BB$.
The latter was found to work well for non-Riemann solver based methods
as well~\cite{vac2}, and is
trivial to carry over to an adaptive grid.
Recently, Janhunen~\cite{janh} and Dellar~\cite{dell} formulated various
arguments to only add the $\nabla \cdot \BB$ related
source term to the induction equation, which restores momentum and energy
conservation. Yet another approach has been forwarded by Marder~\cite{marder},
which relies on diffusing the numerically generated divergence at its
maximal rate. Dedner {\it et al.}~\cite{dedner} revisited this
`parabolic' approach and introduced various variants for hyperbolic
divergence cleaning.
In summary, three `source term' approaches to
handle the $\nabla \cdot \BB$ constraint have respective sources 
$(S_\rho,\SSS_{\rho v},S_e,\SSS_B)$
given by
\begin{equation}
\begin{array}{ccc}
\left(\begin{array}{c}
0 \\
-\divb \BB \\
-\divb \BB\cdot \vv \\
-\divb \vv
\end{array}\right)_{\rm pow} 
 , &
\left(\begin{array}{c}
0 \\
0 \\
0 \\
-\divb \vv
\end{array}\right)_{\rm jan}
 , &
\left(\begin{array}{c}
0 \\
0 \\
\BB \cdot C_{\rm d} \Delta x^2 \nabla \divb \\
C_{\rm d} \Delta x^2 \nabla \divb 
\end{array}\right)_{\rm dif}. 
\end{array}
\label{q-sdivb}
\end{equation}
In this paper, we apply these three simple approaches described
by~(\ref{q-sdivb}) in our
cell-centered magnetic field representation. 
We cross-validate them
on several 2D MHD problems using AMR.
The coefficient $C_{\rm d}=0.2$ in all cases reported.
This diffusive approach could also be altered to keep the energy equation
fully conservative.

\section{Results and efficiency evaluation}\label{s-res}

In this section, we present simulations covering 1D to 3D evolutions.
Our findings are discussed for each problem in what 
follows. 
Refinement is based on the density, if not stated otherwise. 
Tables~\ref{t-time}-\ref{t-time3d}
collect some relevant statistics, including execution times.
All 1D and 2D problems report
timings done on a single 270 MHz MIPS R12000 CPU of an SGI Origin 200.
The 3D cases have been obtained on single or multiple processors of
the SGI Origin 3800, with 500 Mhz MIPS R14000 CPUs.
The efficiency of an AMR
calculation is defined as the ratio between 
its execution time and
the time needed to run the same problem on
a static grid at the resolution of the highest level allowed $l_{\rm max}$. 
We also report on the percentage of computing time spent on
AMR-specific overhead. This overhead includes the total time spent on
the combination of
regridding operations (including the error estimation procedure and the
steps needed for creating and initializing
properly nested new grid hierarchies) and 
update operations, i.e. the replacement of underlying coarse cell values
and the flux fixing of coarse cells neighbouring a finer grid level.
This overhead measure leaves out two factors which are arguably
AMR related,
namely the time spent on storing fluxes across grid boundaries -- which is
negligible -- and the time spent on enforcing boundary conditions
by means of ghost cells. On AMR grid hierarchies, the latter 
involves spatio-temporal interpolation procedures from
underlying coarser levels for filling internal
boundary regions surrounding level $l>1$ grids.

\subsection{1D simulations}\label{ss-res1d}
\subsubsection{Hydrodynamic shock tube}

Our first example considers a Riemann problem for the 1D Euler system
from Harten~\cite{harten}. With a ratio of specific heats $\gamma=1.4$,
the left and right states are
\[ \begin{array}{cclcr}
  (\rho,\rho v_x, e)_L & = & (0.445, & 0.311, & 8.928)
\mathrm{~~~~for~} x<8 \nonumber \\
  (\rho,\rho v_x, e)_R & = & (0.5,   & 0,     & 1.4275)
\mathrm{~~~~for~} x\ge8 \nonumber
\end{array}
\]
in the $x\in[0,14]$ domain with open (zero gradient) boundaries on both sides.
Using 140 cells at level $1$ and allowing for 5 levels with refinement
ratios $r_l=2$, we strive to efficiently achieve the accuracy of a 
static grid simulation exploiting 2240 cells. In Figure~\ref{f-harten}, we
show two solutions at $t=2$: the top panel shows the density as
obtained with FCT. In the bottom panel,
the TVDMU scheme with MC/Woodward limiter was used on all levels.
The extent of the level $l \ge 2$ grids at this time is indicated. 
We set the tolerance $\epsilon_{\rm tol}=
0.005$ and regrid parameters $k=4$, ${\rm E}=0.6$.

As soon as the resulting rarefaction wave, contact discontinuity and shock
are well-separated, higher level grids are only activated around the
contact and the shock discontinuities. Locally, the solutions compare
favorably with static, high-resolution results exploiting the same method:
an equal amount of grid points resolving the contact and shock are found. 
We overplotted the high resolution static grid result in the bottom panel
of Figure~\ref{f-harten}.
Note that the FCT simulation does well
in representing the discontinuities at the cost of introducing spurious 
oscillations, while the TVDMU scheme yields oscillation-free results.
This may well be specific for the ETBFCT variant used here and described
in T\'oth and 
Odstr\v cil~\cite{vac2}. Similar conclusions were reached there
comparing both methods
for static grid simulations and in Steiner {\it et al.}~\cite{oski}
where AMR calculations used FCT. 

The execution times reported in Table~\ref{t-time} are for
simulations up to time $t=2$, using
a fixed time step $\Delta t_{5}=0.000625$. 
They demonstrate an
obtained efficiency of 2.43 when using FCT, which improves to 2.95 when
using TVDLF and up to 3.76 for the TVDMU scheme. This increase in
AMR efficiency is reflected in a decrease of the AMR overhead from
25.7 \% down to 19.8 \%.

\subsubsection{Magnetohydrodynamic Riemann problem}\label{ss-bw}

Our second test problem is the 1.5D MHD Riemann problem introduced by
Brio and Wu~\cite{brio}. With fixed $B_x=0.75$ and $\gamma=2$, the
constant states are
\[ \begin{array}{cclcccr}
  (\rho, \rho v_x, \rho v_y, e, B_y)_L & = & (1, & 0, & 0, & 1.78125, & 1)
\mathrm{~~~~for~~~~} x<0.5 \nonumber \\
  (\rho, \rho v_x, \rho v_y, e, B_y)_R & = & (0.125, & 0, & 0, & 0.88125, & -1)
\mathrm{~~~~for~~~~} x\ge0.5 \nonumber
\end{array}
\]
in the $x\in[0,1]$ domain with open boundaries.
We use 40 base cells and compare two means to achieve locally the resolution
of a 1280 cell static grid. Using the TVDMU scheme and
{\it minmod} limiting on all levels,
we compare in Figure~\ref{f-briowu} the density
at $t=0.1$ from a calculation with $l_{\rm max}=6$ and $r_l=2$
versus one where $l_{\rm max}=4$ and $[r_2,r_3,r_4]=[4,4,2]$. The
tolerance was set to $\epsilon_{\rm tol}=
0.001$ and regrid parameters $k=4$, ${\rm E}=0.8$.
Both solutions accurately capture
the slow compound shock, the contact discontinuity,
and the slow shock. Locally, the obtained solution essentially coincides
with a high resolution static case, again overplotted in the bottom panel.
The timings in Table~\ref{t-time} indicate only little differences in 
efficiency between the AMR simulations exploiting 4 and those exploiting
6 levels. This is true for the TVDLF and the TVDMU scheme. For the
latter discretization, the AMR overhead drops to about 15 \%. 

\subsection{2D simulations}\label{ss-res2d}
\subsubsection{Advection tests}\label{sss-vac2d}

Our first 2D examples consider pure advection problems. On a doubly-periodic
domain $[0,1]\times[0,1]$, a prescribed velocity field $\vv=[v_x,v_y]=[1,1]$ 
diagonally advects a two-dimensional density profile. We perform two tests,
one for a discontinuity dominated profile, and one for a smoothly varying
density pulse.

The discontinuous profile is
a two-dimensional density field representing
the VAC-logo, see Figure~\ref{f-vac}. Within the logo, the density is
set to 2, while it is 0.5 outside. We use a $50\times 50$ level 1 grid,
and allow for three levels with $r_l=2$. The AMR control
parameters are taken $k=4$, ${\rm E}=0.8$,
and $\epsilon_{\rm tol}=0.01$.

Figure~\ref{f-vac} shows the
initial $t=0$ grid structure, along with two snapshots 
at times $t=0.6$ and $t=2$, time-advanced with TVDLF and
a MC/Woodward slope limiter.
To demonstrate the maintained accuracy, cuts
along the $x$-direction at $y=0.5$ from times $t=0, 1, 2$ are
plotted as well. At each integer time, one should regain the $t=0$ situation.
The numerical diffusion immediately smears the discontinuities, but the
resulting density field is in exact agreement with a static grid
result using $200\times 200$ cells. As shown in Table~\ref{t-time}, the
AMR simulation is twice as fast as a static grid run, with only 10 \%
of the execution time spent on regridding operations. 

For the timings reported using the FCT method,  
we used a dimensionally split approach. As pointed out by 
Bell {\it et al.}~\cite{bell}, directional sweeping can be
less favorable to the AMR calculations as additional boundary work
is needed (extra top and bottom layers in the $x$-sweep must
be taken along to start the $y$-sweep). The stencil of the FCT
discretization for directionally unsplit multidimensional simulations requires
filling the corner ghost cells, which we avoid with
the split approach.

Since the exact solution for this advection problem is known, we
illustrate quantitatively that the AMR method reaches the same level
of accuracy as a corresponding high resolution simulation. Table~\ref{t-vac}
summarizes the relative numerical errors obtained
for a variety of AMR runs of different effective resolution.
In general, we define the relative error for a variable $u$ with respect to
a reference solution $u^{\rm ref}$ on a $N\times M$ grid as
\begin{equation}
\delta u =\frac{\sum_{i=1}^N \sum_{j=1}^M \mid u_{i,j} - u^{\rm ref}_{i,j} \mid}
{\sum_{i=1}^N \sum_{j=1}^M \mid u^{\rm ref}_{i,j} \mid} . \label{q-err}
\end{equation}
In this equation, the $u_{i,j}$ values are to be obtained by
extrapolating the AMR solution to a uniform grid of size $N\times M$
corresponding to its (highest level) effective resolution. 
When more than one variable is advanced, the average $\bar{\delta}$ is
taken over all primitive variables $u$. 

Table~\ref{t-vac} demonstrates that (1) the AMR runs achieve the same
(first order) convergence behaviour on this discontinuity dominated
problem as the static grid results; 
(2) the AMR runs are essentially identical to their high resolution
static grid counterparts, with minute improvements noticable when
refining more often (lowering $k$) and/or lowering the tolerance 
$\epsilon_{\rm tol}$. Note that it is advisable to take $k$ consistent
with the buffer zone width $n_{\rm buff}$ used in the regridding
process, as pointed out by Berger and LeVeque~\cite{amrclaw}.
As could be expected, the AMR efficiency is better for more grid levels
than the 3-level one reported in Table~\ref{t-time}: an efficiency of
6 is reached for the $800 \times 800$ AMR run exploiting 5 refinement
levels.

To demonstrate that we obtain second order accuracy on smooth solutions,
we similarly advected a Gaussian bell profile, initially set to
\[ \rho(x,y,t=0)=2 \exp\left(-100\left[(x-0.5)^2+(y-0.5)^2\right]\right) \]
within a radius 0.2 from the center of the square domain and 
constant elsewhere. Again, we compare the obtained numerical solution after
two full advection cycles at time $t=2$ with the exact solution.
Table~\ref{t-bel} shows that we get second order convergence
behaviour to the exact solution, on static and corresponding AMR runs.
The AMR simulations refined every second timestep ($k=2$), while
the tolerance was fixed at $\epsilon_{\rm tol}=0.001$. At this
tolerance, the 5 level AMR result achieving an $800 \times 800$ effective
resolution did not trigger level 5 grids on the entire bell profile. 
Note that we did obtain a solution
with the specified accuracy 0.001, which shows the accuracy
and efficiency of our error estimation procedure.
Hence, this
explains the apparent loss in convergence rate for this AMR run.

Taken together with the 1D test cases from above, these AMR convergence
studies of 2D advection problems demonstrate some general trends. First,
the effective resolution of the AMR grid 
is setting the overall accuracy of the
obtained solution, as desired. Note that it is controlled by the
combination of (1) the base grid resolution; (2) the number of
refinement levels $l_{\rm max}$; and (3) the refinement ratio's in between
consecutive levels $r_l$. 
AMR runs of similar effective resolution
which differ in the exact combinations taken of these three controlling
elements should behave identical (cfr. the test problem shown in 
section~\ref{ss-bw}). These conclusions are valid provided that the
refinement criterion succesfully tracks all flow features. The latter is
influenced by the choice of $\epsilon_{\rm tol}$ and $k$. Table~\ref{t-vac}
demonstrates that only minor differences ensue when varying these
parameters within reasonable bounds. A final set of AMR parameters,
consisting of the efficiencies ${\rm E}$, ${\rm E}_{\rm min}$, 
and $n_{\rm buff}$ plays a role in determining the (highly varying)
number and
the structure of the grids on a certain level. There seems to be little reason
to alter their values dramatically from run to run.

\subsubsection{Double Mach Reflection of a Shock}

A well-known shock dominated hydrodynamical test is the double Mach reflection
of a shock on a wedge, introduced by
Woodward and Colella~\cite{wood}.
On a domain $[0,4]\times[0,1]$, a planar Mach 10 shock
with post- and pre-shock states given by
\[ \begin{array}{cclccr}
  (\rho,\rho v_x, \rho v_y, e)_{\rm post} & = & (8, & 8 \times 8.25\sin 60^\circ, & -8\times 8.25\cos 60^\circ, &  563.5) \nonumber \\
  (\rho,\rho v_x, \rho v_y, e)_{\rm pre} & = & (1.4, & 0, & 0, & 2.5). \nonumber 
\end{array}
\]
makes an angle of $60^\circ$ with the wedge, which is represented as
a reflective boundary located at $x\in[1/6,4]$ and $y=0$.
The adiabatic index is $\gamma=1.4$. A self-similar
pattern develops as the shock reflects off the wedge.
The boundary conditions are fixed to the post-shock state at
the left edge, open at the right edge, a partly fixed, partly reflective
bottom, and a time- and space-dependent boundary value prescription
along the top.

In the AMR simulation, we use TVDLF on a maximum of 4 levels, 
with an $80\times 20$ base resolution and refinement ratios
$r_l=2$. Furthermore, we took $k=6$, ${\rm E}=0.8$,
and a tolerance of $\epsilon_{\rm tol}=0.01$ in the regrid process.
Figure~\ref{f-wc} shows the
density structure at time $t=0.2$. 
We had to use a Courant number of 0.4 for the TVDLF discretization, 
since the $x=1/6$ bottom boundary transition was not
coincident with a cell edge at this resolution. As a result,
a TVDMU simulation developed a carbuncle-type error at the location of the
leading Mach shock intersecting with the wall. A parametric study of
the AMR efficiency in terms of both execution time
and memory needs for a large variety of effective grid resolutions is
reported in Nool and Keppens~\cite{nool}.

Our timings indicate a significant gain
in execution times -- by a factor of 11 --
between the corresponding $640\times 160$ static
grid simulation and the adaptive one. This high efficiency reflects the fact
that the AMR process nicely succeeds
in tracing the localized fronts, so that
the domain coverage by level $l>1$ grids
remains low. In particular, level 4 grids initially cover only 5 \% of the
entire computational domain, with a modest increase to 15 \% at the
end of the calculation at $t=0.2$.
When raising the base resolution to $160 \times 40$ and again
allowing for 4 levels with otherwise identical parameter settings, we reach 
an efficiency of 19.8 with still only 8\% computing time spent on regridding.

\subsubsection{Hydrodynamic Rayleigh-Taylor Instability}\label{ss-hd-rt}

While the previous example represents a typical usage of AMR in hydrodynamical
shock governed evolutions, we now turn to a simulation where a transition
to a fairly global, turbulent state takes place. In such cases, it is
not a priori evident that we can benefit from the use of dynamically triggered
mesh refinement. 

We simulate a 2D Rayleigh-Taylor instability
on a $[0,1]\times [0,1]$ domain, with gravity $\ggg=-\hat{e}_y$ pointing 
downwards. As in Keppens and T\'oth~\cite{vecpar}, we start from a dense
fluid with $\rho_{\rm dense}=1$ above the interface
$y_{\rm int}=0.8+0.05\sin 2\pi x$,
resting on top of a light fluid with $\rho_{\rm light}=0.1$. Both fluids are
at rest at $t=0$ and $\gamma=5/3$. The pressure field is set from a centered
differenced hydrostatic balance $dp/dy=-\rho$, ensuring that 
the pressure about $y_{\rm int}$ equals unity. This configuration is inherently
unstable, while linear stability analysis in incompressible hydrodynamics
predicts the shortest wavelengths growing the fastest. The staircased 
representation of the interface $y_{\rm int}$ initiates such short wavelength
perturbations, while its sinusoidal variation and the periodic side
boundaries select a preferential longer wavelength variation. 
Top and bottom are impenetrable to the fluids.

In~\cite{vecpar}, the fixed $100\times 100$ resolution
and the (small) numerical diffusion associated with the TVDMU scheme 
suppressed the development of much of the small-scale structure. 
Here, we repeat that simulation using AMR to locally achieve a resolution
of $800 \times 800$. As we anticipate a
need for higher resolution spreading
across the entire computational domain, we want to make the cost associated
with all intermediate levels $l\in[2,l_{\rm max}-1]$ as low as possible,
without sacrificing accuracy on the highest level $l_{\rm max}$. 
This can be done by using a level-dependent discretization scheme, such as the
cheap but robust TVDLF method on all levels $l \leq l_{\rm max}-1$, in
combination with a Riemann-solver based method, here TVDMU, on the
highest level $l_{\rm max}=5$ only.
The base grid is $50\times 50$ and all refine ratios
are $r_l=2$. Other AMR parameters are as in the 
Woodward and Colella simulation. 
The snapshot of the logarithm of the density at time
$t=1.9$ shown in Figure~\ref{f-rt} clearly demonstrates that we
reached a very high effective Reynolds number regime.

In Figure~\ref{f-rtc} (left panel)
we plot the domain coverage for each grid level in the
Rayleigh-Taylor simulation as
a function of time. This figure confirms the visual impression of
fine structure spreading across much of the computational volume, such that at
time $t=1.9$, more than 50 \% of the area is at the highest resolution allowed.
Also, the level $l=2$ grids cover the full domain at this time,
making the $l=1$ grid calculation completely obsolete. 
Clearly, since the physical process cascades to
smaller scales, the strategy of minimizing computational cost on lower
lying levels through the use of a cheaper discretization pays off.
In Table~\ref{t-time}, we give an estimate of the
execution time needed for a static $800\times 800$ calculation
from performing 20 time steps at this resolution. As the CFL-limited time step
is observed to decrease monotonically in the AMR run, this allows to deduce both
a lower and upper bound, and we report a mean value in the table.
The corresponding efficiency of the AMR calculation is about 8.3. 
Note also the low value of 3.9 \% of adaptive grid
related overhead.

\subsubsection{Orszag-Tang Vortex System}

The compressible evolution of the Orszag-Tang vortex system~\cite{ot,dahl} is
a well-studied model problem where a shock-governed transition to an MHD
turbulent state takes place. On a doubly periodic domain of 
size $[0,2\pi]\times [0,2\pi]$, velocity ``convection cells''
described by $\vv=[-\sin y,
\sin x]$ are out of phase with magnetic islands resulting from
$\BB=[-\sin y, \sin 2x]$. At time $t=0$, the density and pressure
take the uniform values $\rho=25/9$ and $p=5/3$ with $\gamma=5/3$.  
We evolve this system till time $t=3.14$, using four
levels with a $50\times 50$ resolution at level $1$. With $r_l=2$,
we aim to find the `reference' high resolution $400\times 400$ solution shown
in Fig.~16 from~\cite{divb}. As in that paper, 
our Figure~\ref{f-ot} shows the temperature distribution,
and the fine details agree closely. 
Other parameters used in the AMR simulation are 
$k=6$, ${\rm E}=0.7$, and $\epsilon_{\rm tol}=0.025$. 

To demonstrate the advantages of the mixed strategy where TVDMU is used
on the higher levels only, we report in Table~\ref{t-time} timings for
3 AMR simulations with $l_{\rm max}=4$: one where we use TVDMU
on all levels, one where we use it on levels $l=3$ and $l=4$, mixed with
TVDLF on $l=1,2$, and the advocated choice where TVDMU is active on
level $l=4$ alone. The efficiency increases from 5.85, to 6.28, to 7.38,
respectively. In the right panel of Figure~\ref{f-rtc}, the coverage
of the domain by the four grid levels is shown for the most
efficient case. In agreement with
the initially smooth evolution of the density (uniform at $t=0$), 
higher level grids are only
activated from about $t=0.5$ onwards.
This represents an immediate gain in
the obtained efficiency. As in the Rayleigh-Taylor simulation from above,
the level $l=2$ grids
cover up the entire domain after some time. At the end of the simulation,
this is about to happen for level $3$ as well: a restart with a higher base
level resolution should then be performed.

This 2D MHD problem is also an excellent test for comparing the three
different strategies (referred to as `Powell', `Janhunen' or `diffusive'
and defined by their respective sources in equation~\ref{q-sdivb})
for controlling $\nabla \cdot \BB$ in AMR simulations.
Since the exact solution to this problem is not available analytically,
our best reference is to use the same high resolution solution
used in T\'oth~\cite{divb} for quantitative comparisons. As described in
that paper, using a dimensionally split one-step TVD scheme with MC
limiting, different approaches for controlling $\nabla
\cdot \BB$ still differ slightly at resolutions of $400 \times 400$.
Static grid MHD simulations exploiting a projection scheme~\cite{brack} were
found to yield almost always the most accurate solutions.
Repeating the Orszag-Tang simulation with 4 AMR levels all exploiting
the dimensionally split TVD scheme (for a fair comparison), the obtained
relative errors as defined above at time $t=3.14$
with respect to a reference projection scheme solution are
\begin{equation}
\begin{array}{ccccc}
\bar{\delta}_{\rm pow} = 0.0546 & &
\bar{\delta}_{\rm jan} = 0.0808 & &
\bar{\delta}_{\rm dif} = 0.0465.
\end{array}
\nonumber
\end{equation}
These values should be contrasted to remaining differences among
two high resolution static grid simulations: between 
a constrained transport solution and
a projection scheme approach, a relative error of 
$\bar{\delta} = 0.021$ could be measured. Again, by lowering the tolerance
$\epsilon_{\rm tol}$ to 0.01, improvements up to
$\bar{\delta}_{\rm dif} = 0.02768$ can be obtained. 
These values for the relative errors are also consistent with
the values reported in Table~7 of T\'oth~\cite{divb}. Given the fact
that for a considerable amount of time (for times $t < 1$), 
the AMR simulations could suffice with lower effective resolutions, 
the obtained relative errors are truly optimal.

\subsubsection{Rotor}

A second ideal MHD problem frequently used for comparing static
grid MHD simulations is the rotor problem introduced by Balsara and
Spicer~\cite{bs}. We perform the `first' rotor test as mentioned
in T\'oth~\cite{divb} where the setup is as follows.
On the domain $[0,1]\times[0,1]$, thermal pressure at $t=0$ is
$p=1$, and a uniform magnetic field has $B_x=5/\sqrt{4\pi}$ and $B_y=0$.
We set $\gamma=1.4$, and create a dense disk within $r\equiv
\left[(x-0.5)^2+(y-0.5)^2\right]^{1/2} <r_0=0.1$
with $\rho=10$ and $v_x=-2(y-0.5)/r_0$, $v_y=2(x-0.5)/r_0$. For
$r>r_1=0.115$ the fluid is at rest with $\rho=1$. In between $r_0<r<r_1$,
linear profiles for density and angular speed connect both states.
We run the simulation till $t=0.15$. We strive for a $400 \times 400$
resolution, and use AMR exploiting 4 levels with
$k=6$, $E=0.7$ and $\epsilon_{\rm tol}=0.025$. Flagging is done
on density and both magnetic field components with 
weighting factors 0.8, 0.1 and 0.1, respectively. 
A snapshot of the thermal pressure is shown in Figure~\ref{f-ot}.

Since all important features are captured on the highest grid level
at all times, adaptive simulations with `source term'
treatments to keep the magnetic field solenoidal
yield effectively similar solutions as
static runs. We quantify this latter statement as follows.
We perform several AMR runs with a base grid of $50 \times 50$ and a
varying number of grid levels $l_{\rm max}=2,3,4$. In each case, we use 
TVDLF on the lowest level(s), with TVDMU and MC
limiting on the highest level. The latter scheme was used for the 
reference $400^2$ solution of Figure~18 from~\cite{divb}. This static
grid solution
again used a projection scheme for $\nabla \cdot \BB$. As in the previous
test case, even static grid results yield mutual differences of order
$\bar{\delta}=0.013$ for a flux constrained transport versus a
projection algorithm at this high resolution. 
When we then calculate the relative errors
for AMR runs with effective resolutions increasing from
$100^2$ to $400^2$, these turn out to be
\begin{equation}
\begin{array}{ccccccc}
l_{\rm max} = 2 & &
\bar{\delta}_{\rm pow} = 0.0760 & &
\bar{\delta}_{\rm jan} = 0.0794 & &
\bar{\delta}_{\rm dif} = 0.0753.
\\
l_{\rm max} = 3 & &
\bar{\delta}_{\rm pow} = 0.0358 & &
\bar{\delta}_{\rm jan} = 0.0412 & &
\bar{\delta}_{\rm dif} = 0.0373.
\\
l_{\rm max} = 4 & &
\bar{\delta}_{\rm pow} = 0.0157 & &
\bar{\delta}_{\rm jan} = 0.0248 & &
\bar{\delta}_{\rm dif} = 0.0107.
\end{array}
\nonumber
\end{equation}
These experiments demonstrate that each source term approach 
achieves first order convergence behaviour, as expected for this discontinuity
dominated problem. Furthermore, the errors at effective resolutions of
$400^2$ are comparable to remaining differences between static grid solutions.
As a final note, the advocated mixed discretization approach does not
significantly sacrifice accuracy by the use of the more diffusive scheme
on lower levels. Indeed, when we repeat the simulation with $l_{\rm max}=4$
and exploit TVDMU on all levels, the relative error for a diffusive
approach becomes $\bar{\delta}_{\rm dif} = 0.0096$. 
From this and the previous test case, it seems that the
diffusive approach is preferable over both other source term treatments.

\subsubsection{MHD Kelvin-Helmholtz Instability}

Since the Orszag-Tang and the rotor
system represented ideal MHD evolutions, we include
a resistive MHD simulation taken from Keppens {\it et al.}~\cite{kh2d}.
The initial condition has a uniform density $\rho=1$ and pressure $p=1$,
a sheared and perturbed velocity field given by
\[ \vv=[0.645 \tanh 20 y, 0.01 \sin (2\pi x) \exp(-25 y^2)] \] and a
discontinuously
varying magnetic field $\BB=\pm 0.129 \hat{e}_x$. The magnetic field is thus
of uniform strength, but changes sign at $y=0$ in the
domain $[0,1]\times[-1,1]$. With a resistivity $\eta=10^{-5}$, high resolution
simulations on static $400\times 800$ grids demonstrated in~\cite{kh2d} 
how the nonlinear evolution is characterized by Kelvin-Helmholtz roll-up
triggering tearing behavior. As the Kelvin-Helmholtz instability warps and
amplifies the initial current sheet at $y=0$, several magnetic islands
develop due to local reconnection events, which are clearly visible in the
density pattern (see Fig.~9 in reference~\cite{kh2d}).
Shown in Figure~\ref{f-kht} is a snapshot of the density pattern at $t=4.4$
as obtained with an AMR simulation using 4 levels. Since it is vital
to capture the initial dynamics at the $y=0$ interface accurately, we base
the refinement criterion on the horizontal field component $B_x$ so
that all levels are activated at $t=0$. Other than that,
parameters are
$k=6$, ${\rm E}=0.8$, $\epsilon_{\rm tol}=0.01$. 
Instead of
the one-step TVD method used in~\cite{kh2d}, here we use the combined
TVDLF (on levels $l\leq 3$) and TVDMU (on $l=4$) approach.
The timing reported in Table~\ref{t-time} is for a simulation up to $t=5$,
at which time the transition to a turbulent state has occured. Note the very
high efficiency of 13.9
and the marginal regridding overhead. We used the Powell source terms
in these calculations.

\subsection{3D Simulations}\label{ss-res3d}

\subsubsection{3D Advection}

We made an efficiency assessment of a 3D advection problem by diagonally
advecting a sphere of radius 0.2 in the unit cube where the density was
set to 2, while it was 0.5 external to the sphere. With triple
periodic boundaries, the velocity $\vv=(1,1,1)$ brings the sphere
back to the center at $t=1$. Similar to the 2D advection problem 
evaluated in Table~\ref{t-vac}, accuracy and convergence properties
are equivalent to corresponding high resolution simulations. As reported
in Table~\ref{t-time3d}, a case where AMR exploits 3 levels, with a $40^3$
base grid and refine ratios 2 and 4, demonstrates an efficiency of 19.9.
The AMR overhead has dropped to a rather negligible 4.14 \%.
This problem was run on a single CPU of the SGI Origin 3800.

\subsubsection{3D MHD problems}

As a first 3D MHD simulation, we recovered the
evolution of a magnetized Rayleigh-Taylor instability as in  
Keppens and T\'oth~\cite{vecpar}. 
3D AMR MHD simulations recovering previous high resolution
Kelvin-Helmholtz unstable jet evolutions are documented in
Nool and Keppens~\cite{nool}. In those simulations, 
the diffusive source terms for controlling $\nabla \cdot \BB$ were exploited.
The problem setup for the Rayleigh-Taylor study straightforwardly 
augments the 2D HD simulation in section \ref{ss-hd-rt} with a uniform
horizontal magnetic field $\BB = 0.1 \hat{e}_x$
and a 3D displacement of the
interface separating the heavy from the light fluid according to
$y_{\rm int}=0.8+0.05\sin 2\pi x \sin 2\pi z$.  
The magnetic field suppresses the formation of much of the fine scale
structure. Here, the magnetic field constraint is handled by the Powell source
method.

Using AMR settings with $k=6$, 
${\rm E}=0.7$, $\epsilon_{\rm tol}=0.05$,
the timings 
for two AMR simulations with $l_{\rm max}=2$ and $l_{\rm max}=3$
recovering a $80\times 160 \times 80$ static grid case are 
reported in Table~\ref{t-time3d}.
In both simulations, TVDMU is
used on the highest level only, and a roughly fivefold efficiency is reached. 
Since the base grid is only
$20\times 40 \times 20$,
the problem setup immediately triggers
level $2$ grids on 20 \% and $3$ grids on 10 \% of the total
domain. These values increase to 50 and 20 \%, respectively, at the end
of the calculation. Using the average coverage values of 15 \% and 35 \%,
and taking account of the relative refinement ratios, we can thus expect
that the CPU time is at best reduced to $15 + 35/16 + 100/256 = 17.57$ \%,
giving an optimal efficiency of 5.7, to be contrasted with the
fivefold efficiency actually obtained.  
Since the runtime is dominated by the two
highest levels, the obtained efficiency is truly optimal.
The time spent on the regridding process is fully negligible: less than
2 \% overhead was observed for this problem.
Since one would typically use a higher base grid resolution, much better
efficiencies can be expected to hold in practical 3D MHD simulations.
Indeed, Ziegler~\cite{ziegler} demonstrated an efficiency of 45 on a
purely kinematic 3D magnetic `reconnection' problem (where
only the induction equation~(\ref{q-induction}) was solved)
starting from a base grid
of $100 \times 120 \times 60$ while allowing for 3 grid levels. 

To demonstrate improved efficiencies at higher base grid resolutions, we
set up a 3D MHD implosion problem on the cube $[-0.5,0.5]^3$ with base
grid resolution $60^3$. We again only allow 3 grid levels, with $r_l=2$, so
that an effective resolution of $240^3$ is realized. At time $t=0$,
the problem setup has a uniform magnetic field $\BB=\sqrt{3/5}\hat{e}_y$
pervading a static medium where the pressure is 0.6 external to
the sphere $x^2+y^2+z^2=0.2^2$ and 0.06 inside this sphere. 
The density is $\rho=1$ except for a small off-center sphere within the
low pressure zone: where $(x+0.1)^2+y^2+z^2<0.04^2$ the density is
increased tenfold. We run this simulation using triple periodic boundaries
up to time $t=0.4$,
corresponding to 125 level 1 discrete time steps $\Delta t_1$. 
The solution at this time is shown in Figure~\ref{f-impl},
showing several quantities in cross-cuts at $x=0$, $y=0$, and $z=0$. 
We used TVDLF on all levels and the diffusive source term treatment for
the solenoidal constraint. Other AMR parameters were
$k=2$, ${\rm E}=0.8$, $\epsilon_{\rm tol}=0.01$. This simulation was run
on 16 processors of the SGI Origin 3800 and exploited {\tt OpenMP}
parallelization as explained in reference~\cite{iccs02}. 

During the entire run, the level 3 grids covered on average 7.3 \% of
the entire domain, reaching a maximum of 10.2 \%, while
the level $l=2$ covered an average of 17.8 \%, and a maximal
coverage of 26.6 \%. This already implies
that this AMR run can at best achieve efficiencies in the range $8-11$. 
For a fair
comparison, we measured the wall clock time needed for performing
one timestep at a fixed $240^3$ resolution, while again exploiting
16 processors and using AMRVAC as a domain decompositioner. This high
resolution static grid run needed then 78.9 seconds per time step. 
The average wall clock time needed for setting one $\Delta t_1$ in the
AMR calculation is 37.8 seconds. Taking into account the 
factor 4 in refinement from level 1 to level 3, the obtained efficiency is
8.3, exactly in its optimal range and indeed improved from the fivefold
efficiency in the low base grid Rayleigh-Taylor problem. 
It is also worthwhile mentioning that the memory requirements for
the 3-level AMR versus the static grid run are reduced by a factor of 6. 
Similarly, the space needed for storing one snapshot (in binary format where
the three coordinates plus 8 unknowns per cell are saved in double
precision) is a factor of 9 smaller in the AMR calculation. This also
improves the time needed for IO operations: saving one snapshot
from the AMR run on average took about 8 seconds, while
that took roughly 51 seconds in the high resolution run.
Hence, the overall advantages of performing grid-adaptive
3D MHD simulations are found in optimally reduced CPU times, drastically
reduced memory requirements, and significantly smaller data volumes for
eventual visualization purposes.

\section{Overview and outlook}\label{s-out}

A dimension-independent implementation of the AMR algorithm, in combination
with the modular structure of the Versatile Advection Code, has allowed us
to assess the gain in computing time for a variety of 1D, 2D and 3D 
hydrodynamic and MHD problems. 
The overhead associated with the automated
regridding process is fully negligible for realistic 2D and 3D problems.
A tenfold efficiency is typical for the 2D cases shown, while the
3D timings indicated that very high efficiencies are indeed achievable:
all 3D test problems demonstrated that the optimal efficiency reachable
was in fact obtained. 
This will be exploited in future AMR studies of correspondingly
high resolution 3D MHD simulations of
different, interacting fundamental plasma instabilities and stellar wind
studies. It will be computationally beneficial
to use Riemann solver based methods
like TVDMU on the highest level only, in combination with the robust,
more diffusive, but sufficiently accurate TVDLF discretization on lower levels.

The AMR algorithm introduces a fair amount of control parameters, which
do affect run strategies and the obtainable efficiency. As a general
rule, two important parameter sets can be distinguished: (1) those
controlling the effective resolution (level $l=1$ grid resolution, maximal
allowed levels $l_{\rm max}$, and consecutive refinement ratios $r_l$);
and (2) those determining the refinement criterion (the choice of
variable(s) to refine on, the number $k$ of time steps taken before regridding,
and the tolerance in the criterion $\epsilon_{\rm tol}$). The physics
problem at hand usually suggests suitable variables to use in
the latter. Refining every second timestep $k=2$ is a safe strategy,
while the range for $\epsilon_{\rm tol}$ in all problems studied
above was $\epsilon_{\rm tol} \in [0.001,0.05]$, where the largest values
were used in the higher dimensional problems. As expected, the
effective resolution plays a decisive role in the eventual
efficiency. In cases where HD or MHD instabilities are simulated, their
length scales will put constraints on the base grid resolution to use:
it must be sufficiently high to allow for mode development. The localized
nature of the problem will then help to determine what is a
sufficient number of AMR levels $l_{\rm max}$: this may require some trial
and error while analyzing the first few time steps to find an optimal value. 
Generally, raising the number of levels will beneficially
influence the efficiency.

For multi-dimensional MHD simulations exploiting AMR, three `source term'
approaches listed in equation~(\ref{q-sdivb})
for controlling $\nabla \cdot \BB$ to truncation error were
investigated. 
Quantitative comparisons with reference high resolution simulations showed
that all three work effectively for cell-centered $\BB$ representations.
For the tests included here, the diffusive approach was found to yield the
smallest relative errors. As previously demonstrated by T\'oth~\cite{divb}
for static grid MHD simulations, we conclude that enforcing the solenoidal
character of the magnetic field to machine precision
in a particularly favoured discretization is not an absolute
necessity for obtaining accurate multi-D MHD simulations exploiting AMR.
The simple source term treatments offer a viable alternative to
staggered field representations with considerable complications associated
with constrained transport implementations on AMR grid 
hierarchies~\cite{balsara}. Projection scheme approaches would work
accurately and reliably for AMR MHD simulations as well, but can be expected
to be more complicated algorithmically and computationally costly.

\begin{ack}
VAC is one of the main products of the `Parallel Computational Magneto-Fluid
Dynamics' project, funded by the NWO priority program on Massive Parallel
Computing. This work is part of the research program of the association
agreement of Euratom and `Stichting voor Fundamenteel Onderzoek der Materie'
(FOM) with financial support from NWO and Euratom.
MN benefitted from financial support
from the `Stichting Nationale Computerfaciliteiten' (NCF, Grant NRG 98.10),
also acknowledged for providing computing facilities.
GT has been partly supported by the Hungarian Science Foundation
(OTKA, grant No. T037548).
\end{ack}

\newpage
\begin{table}[h]
\caption{Convergence study for the 2D advection of the VAC logo. The resolution indicates the effective resolution for AMR runs, refining every $k$
(sub-)time steps.} \label{t-vac}
\begin{tabular*}{\textwidth}{@{\extracolsep{\fill}} l l c c r r }

\vphantom{\LARGE B}
resolution & $l_{\rm max}$ & $k$ & $\epsilon_{\rm tol}$  & $\delta_{\rm static}$ & $\delta_{\rm exact}$\\
\hline
$100^2$ & 1 & -- & -- & 0 & 0.12866 \\
        & 2 & 4  & 0.01 & 0.000179 & 0.12867 \\
        & 2 & 2 & 0.01 & 0.000215  & 0.12867 \\
        & 2 & 2 & 0.005 & 0.000132 & 0.12866 \\
\hline
$200^2$ & 1 & -- & -- & 0 & 0.07646 \\
        & 3 & 4 & 0.01 & 0.000301 & 0.07656 \\
        & 3 & 2 & 0.01 & 0.000172 & 0.07648 \\
        & 3 & 2 & 0.005 & 0.000102 & 0.07648 \\
\hline
$400^2$ & 1 & -- & -- & 0 & 0.04612 \\
        & 4 & 4  & 0.01 & 0.001018 & 0.04676 \\
        & 4 & 2 & 0.01 & 0.000222 & 0.04618 \\
        & 4 & 2 & 0.005 & 0.000129 & 0.04617 \\
\hline
$800^2$ & 1 & -- & -- & 0 & 0.02781 \\
        & 5 & 2 & 0.01 & 0.000506 & 0.02811 \\
\hline
\end{tabular*}
\end{table}

\newpage
\begin{table}[h]
\caption{Convergence study for the 2D advection of the Gaussian bell profile.
The resolution indicates the effective resolution for AMR runs 
($l_{\rm max}>1$), 
refining every $k=2$ (sub-)time steps, at a tolerance $\epsilon_{\rm tol}=0.001$.} \label{t-bel}
\begin{tabular*}{\textwidth}{@{\extracolsep{\fill}} l l r }

\vphantom{\LARGE B}
resolution & $l_{\rm max}$ & $\delta_{\rm exact}$\\
\hline
$100^2$ & 1 & 0.0238102 \\
        & 2 & 0.0238104 \\
\hline
$200^2$ & 1 & 0.0058374 \\
        & 3 & 0.0058379 \\
\hline
$400^2$ & 1 & 0.001548 \\
        & 4 & 0.001571 \\
\hline
$800^2$ & 1 & 0.000422 \\
        & 5 & 0.001271 \\
\hline
\end{tabular*}
\end{table}

\newpage
\begin{table}[h]
\caption{Overview of all 1D and 2D Simulations and Timings (in seconds).} \label{t-time}
\begin{tabular*}{\textwidth}{@{\extracolsep{\fill}} c l c r r r }

\vphantom{\LARGE B}
problem  & method & $l_{\rm max}$  & timing & eff. & AMR\%\\
\hline
\vphantom{\LARGE B}
1D HD & FCT & 1 & 18.35 & -- & -- \\
{\it Harten} &   & 5 & 7.55 & 2.43  & 25.7 \\
$2240 \times [0,2]$ & TVDLF  & 1 & 24.12 & --  & -- \\
 & & 5 & 8.17 & 2.95  & 22.5 \\
 & TVDMU  & 1 & 35.39 & --  & -- \\
 & & 5 & 9.41 & 3.76  & 19.8 \\
\hline
\vphantom{\LARGE B}
1.5D MHD & TVDLF & 1 & 4.36 & -- & -- \\
{\it Brio-Wu} &   & 4 & 3.44 & 1.27  & 17.45 \\
 $1280 \times [0,0.1]$ &   & 6 & 3.28 & 1.33  & 17.79 \\
 & TVDMU  & 1 & 7.38 & --  & -- \\
 & & 4 & 4.02 & 1.84  & 14.30 \\
 & & 6 & 3.83 & 1.93  & 15.65 \\
\hline
\vphantom{\LARGE B}
2D Advection & FCT & 1 & 47.91 & -- & -- \\
{\it VAC-logo} & & 3 & 33.70 & 1.42  & 7.9 \\
$200\times 200 \times [0,2]$ & TVDLF  & 1 & 107.36 & --  & -- \\
 & & 3 & 49.32 & 2.18  & 10.1 \\
\hline
\vphantom{\LARGE B}
2D HD {\it Shock reflection} & TVDLF & 1 & 4827.04 & -- & -- \\
$640\times 160 \times [0,0.2]$ & & 4 & 428.72 & 11.26  & 7.13 \\
$1280 \times 320 \times [0,0.2]$ &  & 1 & 33283.7 & -- & -- \\
                    & & 4 & 1679.3 & 19.82  & 8.08 \\
\hline
\vphantom{\LARGE B}
2D HD {\it Rayleigh-Taylor} & TVDMU & 1 & $303715$ & -- & -- \\
$800\times 800\times [0,1.9]$ & TVDLF-MU (4-1)  & 5 & 36504 & 8.32  & 3.91 \\
\hline
\vphantom{\LARGE B}
2D MHD & TVDMU & 1 & 8624.62 & -- & -- \\
{\it Orszag-Tang} & TVDMU  & 4 & 1474.58 & 5.85  & 3.43 \\
  $400\times 400\times [0,3.1]$ & TVDLF-MU (2-2)  & 4 & 1372.82 & 6.28  & 3.49 \\
  & TVDLF-MU (3-1)  & 4 & 1167.91 & 7.38  & 3.94 \\
\hline
\vphantom{\LARGE B}
2D MHD {\it Kelvin-Helmholtz} & TVDMU & 1 & $173572$ & -- & -- \\
$400\times 800 \times [0,5]$  & TVDLF-MU (3-1) & 4 & 12477 & 13.91 & 2.36 \\
\hline
\end{tabular*}
\end{table}

\newpage
\begin{table}[h]
\caption{Overview of 3D Simulations and Timings (in seconds).
Note that the 3D Implosion problem mentions wall clock times for
{\tt OpenMP} parallelized code running on 16 processors of the
SGI Origin 3800.} \label{t-time3d}
\begin{tabular*}{\textwidth}{@{\extracolsep{\fill}} c l c r r }

\vphantom{\LARGE B}
problem  & method & $l_{\rm max}$  & timing & eff. \\
\hline
\vphantom{\LARGE B}
3D Advection  & TVDLF   & 1 &  172197 & --      \\
$320^3\times [0,1]$       &   & 3 &  8653 & 19.90     \\
\hline
\vphantom{\LARGE B}
3D MHD  & TVDMU   & 1 &  $218650$ & --     \\
{\it Rayleigh-Taylor}       & TVDLF-MU (1-1)   & 2 &  46659 & 4.69     \\
       $80\times 160 \times 80 \times [0,1]$ & TVDLF-MU (2-1)   & 3 &  40726 &
5.37     \\
\hline
3D MHD  {\it Implosion} & TVDLF   & 1 &  $39437$ & --     \\
       $240^3 \times [0,0.4]$ &   & 3 &  4733 & 8.3     \\
\hline
\end{tabular*}
\end{table}

\newpage
\begin{figure}[p]
\FIG{
\begin{center}
\resizebox{0.8\textwidth}{0.45\textheight}
{\includegraphics{keppensFig1.eps}}
\end{center}
}
\caption{A hypothetical sequence of three time steps in an AMR simulation
allowing for 4 nested refinement levels. Vertical arrows indicate a temporal
advance, horizontal arrows correspond to update operations ensuring
conservation and consistency, while the automated regridding operation
is called at the position of the grey circles. Regridding is invoked
when at least $k=2$ time steps are taken.
}
\label{f-time}
\end{figure}

\newpage
\begin{figure}[htbp]
\FIG{
\begin{center}
\resizebox{0.8\textwidth}{!}
{\includegraphics{update.eps}}
\end{center}
}
\caption{Illustration of the update and flux fix step. An underlying
coarse (level $l-1$) cell value is merely replaced by a conservative average. 
A coarse
cell bordering a finer level $l$ grid needs its numerical flux across
the edge $x_{i+\frac{1}{2}}$ replaced by the more accurate fluxes used
on the level $l$.}
\label{f-update}
\end{figure}

\newpage
\begin{figure}[htbp]
\FIG{
\begin{center}
\resizebox{6.2cm}{6.7cm}
{\includegraphics{grid2b.bw.eps}}%
\resizebox{0.5\textwidth}{!}
{\includegraphics{grid3b.bw.eps}}

\resizebox{0.5\textwidth}{!}
{\includegraphics{grid4b.bw.eps}}%
\resizebox{0.5\textwidth}{!}
{\includegraphics{grid5b.bw.eps}}%

\end{center}
}
\caption{The regridding operation illustrated in 2D. {\it Top left:} 5 cells
are flagged for refinement on 2 adjacent level $l$ grids. The edge of the 
computational
domain is indicated by thick solid lines. Each flagged cell is surrounded by a
2-cell buffer zone. {\it Top right:} Buffer cells not located on
level $l$ grids are discarded. A previously formed, enlarged level $l+2$ grid
(small rectangle) is to be projected for proper nesting in the
level $l+1$ grids being formed. {\it Bottom left:} discard cells which would
violate the proper nesting criterion. Note the flagged cell which is lost!
{\it Bottom right:} A possible grid structure for level $l+1$, whereby two
extra level $l$ cells (hashed squares) are taken along.
Note that the level $l+1$ grids may
overlap multiple level $l$ grids.}
\label{f-grid}
\end{figure}

\newpage
\begin{figure}[p]
\FIG{
\begin{center}
\resizebox{\textwidth}{!}
{\includegraphics{keppensFig2a.eps}}

\resizebox{\textwidth}{!}
{\includegraphics{keppensFig2b.eps}}

\end{center}
}
\caption{The density structure at time $t=2$ for the 1D hydrodynamic
Harten shock tube. Using AMR with 5 grid levels, we confront a solution
obtained with FCT (top panel) with a TVDMU result (bottom panel). The
extent of the level $l>1$ grids is indicated by the solid bars.
A high resolution static grid result is overplotted as a solid line
in the bottom panel.}
\label{f-harten}
\end{figure}

\newpage
\begin{figure}[p]
\FIG{
\begin{center}
\resizebox{0.9\textwidth}{0.28\textheight}
{\includegraphics{keppensFig3a.eps}}

\vspace*{-0.5cm}
\resizebox{0.9\textwidth}{0.28\textheight}
{\includegraphics{keppensFig3b.eps}}

\end{center}
}
\vspace*{-0.5cm}
\caption{The density variation at $t=0.1$ in the Brio-Wu shocktube problem.
A solution using 6 refinement levels (top panel) is compared with one
where 4 levels and correspondingly higher refinement ratios were used
(bottom panel). The
extent of the level $l>1$ grids is indicated by the solid bars
and the bottom panel shows a high resolution reference result.}
\label{f-briowu}
\end{figure}

\newpage
\begin{figure}[p]
\FIG{
\begin{center}
\parbox{0.49\textwidth}{
\resizebox{0.5\textwidth}{0.5\textwidth}
{\includegraphics{keppensFig4a.eps}}
}
\parbox{0.49\textwidth}{
\resizebox{0.5\textwidth}{0.166666\textwidth}
{\includegraphics{keppensFig4b.eps}}

\resizebox{0.5\textwidth}{0.166666\textwidth}
{\includegraphics{keppensFig4c.eps}}

\resizebox{0.5\textwidth}{0.166666\textwidth}
{\includegraphics{keppensFig4d.eps}}
}

\resizebox{0.5\textwidth}{0.5\textwidth}
{\includegraphics{keppensFig4e.eps}}%
\resizebox{0.5\textwidth}{0.5\textwidth}
{\includegraphics{keppensFig4f.eps}}
\end{center}
}
\caption{Diagonal advection of the VAC-logo on a doubly periodic domain.
We show the initial 3-level grid structure at top left, and snapshots
at $t=0.6$ (bottom left) and after two full advection cycles (bottom right).
Cuts at $y=0.5$ for times $t=0,1,2$ (top right) demonstrate the
maintained accuracy.}
\label{f-vac}
\end{figure}

\newpage
\begin{figure}[p]
\FIG{
\begin{center}
\resizebox{\textwidth}{!}
{\includegraphics{keppensFig5.eps}}
\end{center}
}
\caption{The density structure at $t=0.2$ for the Woodward and Colella
reflected shock problem, obtained with an AMR simulation using 4 grid levels
and a $80 \times 20$ base grid.}
\label{f-wc}
\end{figure}

\newpage
\begin{figure}[p]
\FIG{
\begin{center}
\resizebox{8cm}{!}
{\includegraphics{keppensFig6.eps}}
\end{center}
}
\caption{The logarithm of the density at time $t=1.9$
shows the fine scale mixing process of a heavy fluid (dark) into a
lighter one underneath under the influence of gravity. Only
the level $l=5$ grids are indicated.}
\label{f-rt}
\end{figure}

\newpage
\begin{figure}[p]
\FIG{
\begin{center}
\resizebox{0.5\textwidth}{!}
{\includegraphics{keppensFig7a.eps}}%
\resizebox{0.5\textwidth}{!}
{\includegraphics{keppensFig7b.eps}}
\end{center}
}
\caption{For the Rayleigh-Taylor simulation from Figure~\ref{f-rt}
(left) and for the Orszag-Tang vortex simulation from Figure~\ref{f-ot}
(right),
the domain coverage of the grid levels used in the AMR simulation as a
function of time.}
\label{f-rtc}
\end{figure}

\newpage
\begin{figure}[p]
\FIG{
\begin{center}
\resizebox{0.5\textwidth}{!}
{\includegraphics{keppensFig8a.eps}}%
\resizebox{0.5\textwidth}{!}
{\includegraphics{keppensFig8b.eps}}
\end{center}
}
\vspace*{-0.5cm}
\caption{{\it Left:} the temperature at $t=3.14$
for the Orszag-Tang magnetized vortex system as
obtained with a 4 level AMR simulation.
{\it Right:} the thermal pressure at $t=0.15$ for the rotor problem
as obtained with a 4 level AMR simulation.}
\label{f-ot}
\end{figure}

\newpage
\begin{figure}[p]
\FIG{
\begin{center}
\resizebox{\textwidth}{!}
{\includegraphics{keppensFig9.eps}}
\end{center}
}
\vspace*{-0.5cm}
\caption{The density structure at time $t=4.4$ for a Kelvin-Helmholtz
unstable shear flow configuration containing a current sheet shows the
triggering of magnetic islands through tearing instabilities. Only the
central part of the computational domain is shown.}
\label{f-kht}
\end{figure}

\newpage
\begin{figure}[p]
\FIG{
\begin{center}
\resizebox{0.5\textwidth}{!}
{\includegraphics{implmhd33.A.eps}}%
\resizebox{0.5\textwidth}{!}
{\includegraphics{implmhd33.B.eps}}

\resizebox{0.5\textwidth}{!}
{\includegraphics{implmhd33.C.eps}}%
\resizebox{0.5\textwidth}{!}
{\includegraphics{implmhd33.D.eps}}
\end{center}
}
\vspace*{-0.5cm}
\caption{3D MHD implosion problem at time $t=0.4$. {\bf Top left:}
in the plane $z=0$, we show a schlieren plot of the pressure, the magnetic field lines,
and the velocity field. {\bf Top right:} schlieren plot of the density in $y=0$.
{\bf Bottom panels:} density (left) and thermal pressure (right) variation
in the plane $x=0$.}
\label{f-impl}
\end{figure}


\begin{thebibliography}{9}
\bibitem{almgren}
   A.S. Almgren, J.B. Bell, P. Colella, L.H. Howell and M.L. Welcome,
   A conservative Adaptive Projection Method for the Variable Density
 Incompressible Navier-Stokes Equations,
   {\em J. Comput. Phys.} {\bf 142} (1998) 1.
\bibitem{balsara}
   D.S. Balsara,
   Divergence-Free Adaptive Mesh Refinement for Magnetohydrodynamics,
   {\em J. Comput. Phys.} {\bf 174} (2001) 614.
\bibitem{bs}
   D.S. Balsara and D.S. Spicer,
   A staggered mesh algorithm using high order Godunov fluxes to ensure
solenoidal magnetic fields in magnetohydrodynamic simulations,
   {\em J. Comput. Phys.} {\bf 149} (1999) 270.
\bibitem{bell}
   J. Bell, M. Berger, J. Saltzman and M. Welcome,
   Three-dimensional adaptive mesh refinement for hyperbolic conservation
laws,
  {\em SIAM J. Sci. Comp.} {\bf 15} (1994) 127.
\bibitem{bell2}
  J.B. Bell, P. Colella and H.M. Glaz,
  A second-order projection method for the incompressible Navier-Stokes
  equations,
   {\it J. Comput. Phys.} {\bf 85} (1989) 257.
\bibitem{berger1}
   M.J. Berger, 
   Data structures for adaptive grid generation, 
   {\it SIAM J. Sci. Stat. Comput.} {\bf 7} (1986) 904.
\bibitem{berger}
   M.J. Berger and P. Colella, 
   Local adaptive mesh refinement for shock hydrodynamics, 
   {\it J. Comput. Phys.} {\bf 82} (1989) 64.
\bibitem{amrclaw}
   M.J. Berger and R.J. LeVeque, 
   Adaptive mesh refinement using wave-propagation algorithms for 
   hyperbolic systems, 
   {\it SIAM J. Numer. Anal.} {\bf 35} (1998) 2298.
\bibitem{rigout}
   M.J. Berger and I. Rigoutsos,
   An Algorithm for Point Clustering and Grid Generation,
   {\em IEEE Transactions on Systems, Man and Cybernetics} {\bf 21}, 
   (1991) 1278.
\bibitem{lbnl}
   Berkeley Lab AMR homepage at {\tt http://seesar.lbl.gov/}.
\bibitem{boris}
   J. P. Boris and D. L. Book,
   Flux-corrected transport. I. SHASTA, A fluid transport
   algorithm that works,
   {\it J. Comput. Phys.} {\bf 11} (1973) 38.
\bibitem{brack}
   J.U. Brackbill and D.C. Barnes,
   The effect of nonzero $\nabla \cdot \BB$ on the numerical solution of
the magnetohydrodyanmic equations,
   {\it J. Comput. Phys.} {\bf 35} (1980) 426.
\bibitem{brio}
   M. Brio and C.C. Wu, 
   An upwind differencing scheme for the equations of ideal 
   magnetohydrodynamics, 
   {\em J. Comput. Phys.} {\bf 75} (1988) 400.
\bibitem{chiang}
  Y.-L. Chiang, B. van Leer and K.G. Powell, 
  Simulation of unsteady inviscid flow on an adaptively refined Cartesian grid,
  {\em AIAA} {\bf 92-0443} (1992).
\bibitem{dell}
  P.J. Dellar,
  A note on magnetic monopoles and the one-dimensional MHD Riemann problem,
   {\em J. Comput. Phys.} {\bf 172} (2001) 392.
\bibitem{dedner}
  A. Dedner, F. Kemm, D. Kr\"oner, C.-D. Munz, T. Schnitzer and M. Wesenberg,
  Hyperbolic divergence cleaning for the MHD equations,
   {\em J. Comput. Phys.} {\bf 175} (2002) 645.
\bibitem{grauer}
   H. Friedel, R. Grauer and C. Marliani, 
   Adaptive mesh refinement for singular
   current sheets in incompressible magnetohydrodynamic flows,
   {\em J. Comput. Phys.} {\bf 134} (1997) 190.
\bibitem{garcia}
   A.L. Garcia, J.B. Bell, W.Y. Crutchfield and B.J. Alder,
   Adaptive mesh and algortihm refinement using direct simulation Monte Carlo,
   {\em J. Comput. Phys.} {\bf 154} (1999) 134.
\bibitem{harten}
   A. Harten, 
   High resolution schemes for hyperbolic conservation laws,
   {\em J. Comput. Phys.} {\bf 49} (1983) 357.
\bibitem{janh}
  P. Janhunen,
  A positive conservative method for magnetohydrodynamics based on HLL and 
  Roe methods,
  {\em J. Comput. Phys.} {\bf 160} (2000) 649.
\bibitem{amrvac1}
   R. Keppens, M. Nool, P.A. Zegeling and J.P. Goedbloed, 
   Dynamic grid adaptation for computational magnetohydrodynamics, 
   {\em Lecture Notes in Computer Science} {\bf 1823} (2000) 61.
\bibitem{vecpar}
   R. Keppens and G. T\'oth, 
   Simulating magnetized plasma with the Versatile Advection Code, 
   {\em Lecture Notes in Computer Science} {\bf 1573} (1999) 680.
\bibitem{vacpar}
 R. Keppens and G. T\'oth, 
Using high performance fortran for magnetohydrodynamic simulations,
{\it Parallel Computing} {\bf 26} (2000) 705.
\bibitem{impl2}
   R. Keppens, G. T\'oth, M.A. Botchev and A. van der Ploeg, 
   Implicit and semi-implicit schemes: algorithms, 
   {\em Int. J. Numer. Meth. Fluids} {\bf 30} (1999) 335.
\bibitem{kh2d}
   R. Keppens, G. T\'oth, R.H.J. Westermann and J.P. Goedbloed,
   Growth and saturation of the Kelvin-Helmholtz instability with parallel and
   antiparallel magnetic fields, 
   {\it J. Plasma Phys.} {\bf 61} (1999) 1.
\bibitem{parcfd}
   R. Keppens, M. Nool and J.P. Goedbloed,
   Zooming in on 3D magnetized plasmas with grid-adaptive simulations,
   in: Parallel Computational Fluid Dynamics -- Practice and Theory,
   P. Wilders et al. (eds.), Elsevier Science B.V. (2002) 215. 
\bibitem{iccs02}
   R. Keppens and G. T\'oth,
   {\tt OpenMP} parallelism for multi-dimensional grid-adaptive
   magnetohydrodynamic simulations,
   {\em Lect. Notes Comp. Science} {\bf 2329} (2002) 940.
\bibitem{lang}
   J.O. Langseth and R.J. LeVeque, A wave propagation method for
   three-dimensional hyperbolic conservation laws,
   {\em J. Comput. Phys.} {\bf 165} (2000) 126.
\bibitem{webamrclaw}
   R.J. LeVeque, {\tt http://www.amath.washington.edu/}$\sim${\tt claw/},
   CLAWPACK 4.0 website.
\bibitem{leveque}
   R.J. LeVeque, Wave propagation algorithms for multidimensional
 hyperbolic systems, 
   {\em J. Comput. Phys.} {\bf 131} (1997) 327.
\bibitem{paramesh}
  P. MacNeice, K.M. Olson, C. Mobarry, R. de Fainchtein and C. Packer,
 PARAMESH: A parallel adaptive mesh refinement community toolkit,
 {\em Comput. Phys. Commun.} {\bf 126} (2000) 330.
\bibitem{marder}
 B. Marder, A method for incorporating Gauss' law into electromagnetic PIC
  codes,
   {\em J. Comput. Phys.} {\bf 68} (1987) 48.
\bibitem{nool}
 M. Nool and R. Keppens, AMRVAC: a multidimensional grid-adaptive
magnetofluid dynamics code, 
 {\em Comp. Meth. Applied Math.} {\bf 2} (2002) 92.
\bibitem{ot}
   A. Orszag and C.M. Tang, 
   Small-scale structure of two-dimensional magnetohydrodynamic turbulence, 
   {\em J. Fluid Mech. } {\bf 90} (1979) 129.
\bibitem{dahl}
  J.M. Picone and R.B. Dahlburg,
  Evolution of the Orszag-Tang vortex system in a compressible medium. 
  II - Supersonic flow,
  {\em Phys. Fluids B} {\bf 3} (1991) 29.
\bibitem{powell}
   K.G. Powell, 
   An approximate Riemann solver for magnetohydrodynamics 
   (that works in more than one dimension), 
   {\it ICASE Report No 94-24, Langley, VA} (1994).
\bibitem{powel}
   K.G. Powell, P.L. Roe, T.J. Linde, T.I. Gombosi and D.L. De Zeeuw,
   A solution-adaptive upwind scheme for ideal magnetohydrodynamics, 
   {\em J. Comput. Phys.} {\bf 154} (1999) 284.
\bibitem{roe}
   P. L. Roe,
   Approximate Riemann solvers, parameter vectors,
   and difference schemes,
   {\it J. Comput. Phys.} {\bf 43} (1981) 357.
\bibitem{roe-balsara}
   P. L. Roe and D. S. Balsara,
   Notes on the eigensystem of magnetohydrodynamics,
   {\it SIAM J. Appl. Math.} {\bf 56} (1996) 57.
\bibitem{oski}
   O. Steiner, M. Kn\"olker and M. Sch\"ussler, 
   Dynamic interaction of convection with magnetic flux sheets: 
   first results of a new MHD code, in {\em Solar Surface Magnetism}, 
   edited by R.J. Rutten and C.J. Schrijver (Kluwer, Dordrecht, 1993).
\bibitem{strang}
   G. Strang,
   On the construction and comparison of difference schemes,
   {\it SIAM J. Numer. Anal.} {\bf 5} (1968) 506.
\bibitem{divb}
   G. T\'oth, 
   The $\nabla \cdot \BB =0$ constraint in shock-capturing
   magnetohydrodynamic codes, 
   {\em J. Comput. Phys.} {\bf 161} (2000) 605.
\bibitem{lasy}
   G. T\'oth, 
   The LASY preprocessor and its application to general multi-dimensional 
   codes,
   {\em J. Comput. Phys.} {\bf 138} (1997) 981.
\bibitem{vac1}
   G. T\'oth, 
   A general code for modeling MHD flows on parallel computers:
   Versatile Advection Code, 
   {\em Astrophys. Lett. \& Comm.} {\bf 34} (1996) 245. \hfill \\
   See {\tt http://www.phys.uu.nl/}$\sim${\tt toth}.
\bibitem{impl1}
   G. T\'oth, R. Keppens and M.A. Botchev, 
   Implicit and semi-implicit schemes in the Versatile Advection Code: 
   numerical tests, 
   {\em Astron. Astrophys.} {\bf 332} (1998) 1159.
\bibitem{vac2}
   G. T\'oth and D. Odstr\v cil, 
   Comparison of some Flux Corrected Transport and Total Variation Diminishing 
   numerical schemes for hydrodynamic and magnetohydrodynamic problems,
   {\em J. Comput. Phys.} {\bf 128} (1996) 82.
\bibitem{toth-roe}
    G. T\'oth and P. L. Roe,
    Divergence- and Curl-Preserving Prolongation and Restriction Formulas,
    {\em J. Comput. Phys.} {\bf 180} (2002) 736.
\bibitem{vanleer}
   B. van Leer,
   Towards the ultimate conservative difference scheme.
   V. A Second order sequel to Godunov's method,
   {\it J. Comput. Phys.} {\bf 32} (1979) 101.
\bibitem{amaze}
   R. Walder and D. Folini, 
   A-MAZE: A code package to compute 3D magnetic flows,
   3D NLTE radiative transfer, and synthetic spectra, 
   in {\it Thermal and Ionization aspects of flows from hot stars: 
   observations and theory},
   ASP Conference Series {\bf 204} (2000) 281.
\bibitem{walder2}
   R. Walder, D. Folini and S. Motamen, Colliding winds in WR binaries:
   further developments within a complicated story, in {\it
   Proc. IAU symposium No. 193}, eds. K.A. van der Hucht, G. Koenigsberger
   and P.R.J. Eenens (1999) 298.
\bibitem{wood}
   P.R. Woodward and P. Colella, 
   The numerical simulation of two-dimensional fluid flow with strong shocks, 
   {\em J. Comput. Phys.} {\bf 54} (1984) 115.
\bibitem {yee}
   H. C. Yee,
   A class of high-resolution explicit and implicit shock-capturing methods,
   {\it NASA TM-101088} (1989).
\bibitem{ziegler}
   U. Ziegler, 
   A three-dimensional Cartesian adaptive mesh code for compressible 
   magnetohydrodynamics, 
   {\it Comput. Phys. Commun.} {\bf 116} (1999) 65.
\end{thebibliography}
\end{document}